\documentclass[epsfig,12pt]{article}
\usepackage{graphics}
\usepackage{epsfig}
\usepackage{amssymb}
\usepackage{amsmath}
\begin{document}

\baselineskip .7cm

\author{ Navin Khaneja \thanks{To whom correspondence may be addressed. Email:navinkhaneja@gmail.com} \thanks{Department of Electrical Engineering, IIT Bombay, Powai - 400076, India.}
\ Ashutosh Kumar \thanks{Department of Biosciences and Bioengineering , IIT Bombay, Powai- 400076, India.}}

\vskip 4em

\title{\bf Broadband excitation by method of double sweep}

\maketitle

\vskip 3cm

\begin{center} {\bf Abstract} \end{center}
The paper describes the design of broadband excitation pulses in high resolution NMR by method of double sweep.  We first show the design of a pulse sequence that produces broadband excitation to the equator of Bloch sphere with phase linearly dispersed as frequency. We show how this linear dispersion can then be refocused by nesting free evolution between two adiabatic inversions (sweeps). We then show how this construction can be generalized to exciting arbitrary large bandwidths without increasing the peak rf-amplitude and by incorporating more adiabatic sweeps. Finally, we show how the basic design can then be modified to give a broadband $x$ rotation over arbitrary large bandwidth and with limited rf-amplitude. Experimental excitation profiles for the residual HDO signal in a sample of $99.5\%$ D$_2$0 are displayed as a function of resonance offset. Application of the excitation is shown for $^{13}$C excitation in a labelled sample of Alanine. 

\vskip 3em

\section{Introduction}
The excitation pulse is ubiquitous in FT-NMR, being the starting point of all experiments. With increasing
field strengths in high resolution NMR, sensitivity and resolution comes with the challenge of
uniformly exciting larger bandwidths. At a field of 1 GHz, the target bandwidth is $50$ kHz for excitation of entire 200 ppm $^{13}$C chemical shifts. 
The required $25$ kHz hard pulse exceeds the capabilities of most
$^{13}$C  probes and poses additional problems in phasing the spectra. Towards this end, several methods have been developed for 
Broadband excitation/inversion, which have reduced the phase variation of the excited magnetization as a function of the resonance offset. 
These include composite pulses, adiabatic sequences, polycromatic sequences, phase alternating
pulse sequences and optimal control pulse design \cite{comp1}-\cite{bulu}.

In this paper, we propose a new approach for design of broadband excitation and rotation pulses, called the method of double sweep. In this approach, a pulse sequence that produces broadband excitation to equator of Bloch sphere with phase linearly dispersed as frequency is designed. This linear dispersion is then refocused by nesting free evolution between two adiabatic inversions (sweeps). This construction is generalized to exciting arbitrary large bandwidths without increasing the peak rf-amplitude by incorporarting more adiabatic sweeps. Finally, we show how the basic design can then be modified to give a broadband $x$ rotation over arbitrary large bandwidth and limited rf-amplitude.

The paper is organized as follows. In section 2, we present the theory behind double sweep excitation. In section 3, we show simulation results for broadband
excitation and broadband rotation pulses designed using double sweep technique. In section 4, we present
experimental data. Finally we conclude in section 5, with discussion and outlook.

\section{Theory}

We consider the problem of broadband excitation. Consider the evolution of spinor(We use $I_\alpha$ to denote the Pauli matrix such that $\alpha \in \{x, y, z \}$ ) of a spin $\frac{1}{2}$ in a rotating frame, rotating around $z$ axis at Larmor frequency.

\begin{equation}
\label{eq:basiceq}
\frac{d |\psi \rangle }{dt} = -i ( \omega I_z + A(t) \cos \theta(t) I_x + A(t) \sin \theta(t) I_y )  |\psi \rangle,
\end{equation}where $A(t)$ and $\theta(t)$ are amplitude and phase of rf-pulse and we normalize the chemical shift in the range $\omega \in [-1, 1]$. Let $u(t) = A(t) \exp(-j \theta)$.

\begin{equation}
\frac{d |\psi \rangle }{dt} = -\frac{i}{2} \left [ \begin{array}{cc} \omega & u(t)\\ u^\ast(t) & -\omega \end{array} \right ] |\psi \rangle.
\end{equation}

Going into interaction frame of chemical shift, we can write the evolution as

\begin{equation}
\label{eq:solution}
|\psi(T) \rangle = \left [ \begin{array}{cc} \exp(\frac{-i \omega T}{2})  & 0 \\ 0 & \exp(\frac{i \omega T}{2}) \end{array} \right ] \exp (\oint_0^T \frac{-i}{2} \left [ \begin{array}{cc} 0 & u(t) \exp(i \omega t) \\ u^\ast(t) \exp(-i \omega t) & 0 \end{array} \right ]) |\psi(0) \rangle.
\end{equation}

We write,

\begin{equation}
\int_0^T  u(t) \exp(i \omega t) = \exp( \frac{i \omega T}{2}) \int_0^T  u(t) \exp(i \omega (t - \frac{T}{2})).
\end{equation}

We design $u(t)$ such that for all $\omega$ we have

\begin{equation}
 \int_0^T  u(t) \exp(i \omega (t - \frac{T}{2})) \sim \frac{\pi}{2}.
\end{equation}

Divide $[0, T]$ in intervals of step, $\Delta t$, over which $u(t)$ is constant. Call them $\{u_{-M}, \dots, u_{-k}, \dots, u_0 \}$ over
$[0, \frac{T}{2}]$ and  $\{u_{0}, \dots, u_{k}, \dots, u_M \}$ over $[\frac{T}{2}, T]$.

\begin{equation}
\label{eq:fourierapprox}
 \int_0^T  u(t) \exp(i \omega (t - \frac{T}{2})) \sim (u_0 + \sum_{k = -M}^{M} u_k \exp(-i \omega k \Delta t))\Delta t, 
\end{equation}

where write $\Delta t = \frac{\pi}{N}$ and choose $u_k$ real with $u_k = u_{-k}$. Then we get 

\begin{equation}
\label{eq:fourierseries}
 \int_0^T  u(t) \exp(i \omega (t - \frac{T}{2})) \sim 2 \sum_{k = 0}^{M} u_k \cos (\omega k \Delta t) \Delta t =  2 \sum_{k = 0}^{M} u_k \cos (k x) \Delta t,
\end{equation} where for $x \in [ -\frac{\pi}{N}, \frac{\pi}{N} ]$, we have $2 \sum_{k = 0}^{M} u_k \cos (k x) \Delta t \sim \frac{\pi}{2}$ and $0$ for $x$ outside this range. This is a Fourier series, and we get the Fourier coeffecients as, 

\begin{equation}
\label{eq:fouriercoeff}
u_0 = \frac{1}{4}\ ; \ \ u_k = \frac{\sin(\frac{k \pi}{N})}{\frac{2 k \pi}{N}}.
\end{equation}

Approximating,

\begin{eqnarray}
\label{eq:frame1}
\nonumber \exp (\oint_0^T \left [ \begin{array}{cc} 0 & \frac{-i u(t) \exp(i \omega t)}{2} \\ \frac{-i u^\ast(t) \exp(-i \omega t)}{2} & 0 \end{array} \right ]) &\sim&  \exp(  \frac{-i}{2} \left [ \begin{array}{cc} 0 & \int_0^T u(t) \exp(i \omega t) \\ \int_0^T u^\ast(t) \exp(-i \omega t) & 0 \end{array} \right ])\\ \nonumber &\sim& \exp(  \frac{-i}{2} \left [ \begin{array}{cc} 0 & \exp( \frac{i \omega T}{2}) \frac{\pi}{2} \\  \exp( \frac{-i \omega T}{2}) \frac{\pi}{2} & 0 \end{array} \right ]) \\ &=& \exp(-i \frac{\pi}{2} (\cos \frac{\omega t}{2} I_x - \frac{\sin \omega t}{2} I_y)) \\ &=&   \frac{1}{\sqrt{2}} \left [ \begin{array}{cc} 1 & -i \exp( \frac{i \omega T}{2})\\  -i \exp( \frac{-i \omega T}{2}) & 1 \end{array} \right ]).
\end{eqnarray}

Statrting from the initial state $|\psi(0) \rangle = \left [ \begin{array}{c} 1 \\ 0 \end{array} \right ]$, we have from Eq. \ref{eq:solution}, 

\begin{equation}
|\psi(T) \rangle = \frac{1}{\sqrt 2} \left [ \begin{array}{c} \exp( \frac{-i \omega T}{2}) \\ -i  \end{array} \right ].
\end{equation}

This state is dephased on the Bloch sphere equator. We show how using a double adiabatic sweep, we can refoucs the phase. Let $\Theta(\omega)$ be the rotation for a adiabatic inversion of a spin. We can use Euler angle decomposition to write, 

\begin{equation}
\Theta(\omega) = \exp(-i \alpha(\omega) I_z) \exp(-i \pi I_x) \exp(-i \beta(\omega) I_z).
\end{equation} The center rotation shold be $\pi$ for $\Theta(\omega)$ to do inversion of $I_z \rightarrow -I_z$.

We can use this to refocus the forward free evolution. Observe

\begin{equation}
\Delta (\omega, \frac{T}{2}) = \exp(\frac{i \omega T}{2} I_z) = \Theta(\omega) \exp(\frac{-i \omega T}{2} I_z) \Theta(\omega).
\end{equation}

Then

\begin{equation}
\label{eq:firstexcitation}
\Theta(\omega) \ \exp(\frac{-i \omega T}{2} I_z) \ \Theta(\omega) \ \frac{1}{\sqrt 2} \left [ \begin{array}{c} \exp( \frac{-i \omega T}{2}) \\ -i  \end{array} \right ] = \frac{\exp( \frac{-i \omega T}{4})}{\sqrt 2} \left [ \begin{array}{c} 1 \\ -i  \end{array} \right ],
\end{equation} which is a broadband excitation.

\begin{figure}
\centering
\includegraphics[scale=.5]{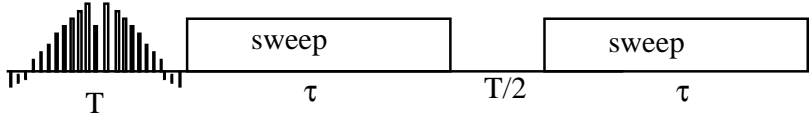}
\caption{The figure shows the basic pulse sequence with a double sweep that performs broadband excitation as in Eq. (\ref{eq:firstexcitation}).}\label{fig:pulsesequence}
\end{figure}

The pulse sequence consists of a sequence of x-phase pulses, which produce the evolution

\begin{equation}
U(\omega, \theta) = \left [ \begin{array}{cc} \exp(\frac{-i \omega T}{2})  & 0 \\ 0 & \exp(\frac{i \omega T}{2}) \end{array} \right ]\exp(  \frac{-i}{2} \left [ \begin{array}{cc} 0 & \exp( \frac{i \omega T}{2}) \theta \\  \exp( \frac{-i \omega T}{2}) \theta & 0 \end{array} \right ]),
\end{equation} where $\theta = \frac{\pi}{2}$, as described above, followed by a double sweep rotation $\Delta (\omega, \frac{T}{2})$. This required a 
peak amplitude of $u(t) \sim \frac{1}{2}$. If peak amplitude is $ \frac{1}{4} \leq u < \frac{1}{2}$, we can prepare $U(\omega, \frac{\pi}{4})$ and combine two such rotations with two double sweeps as follows

\begin{equation}
\label{eq:secondexcitation}
U_1 = \Delta (\omega, \frac{T}{2}) \ U(\omega, \frac{\pi}{4}) \ \Delta (\omega, T) \ U(\omega, \frac{\pi}{4}).
\end{equation}

In general, if $ \frac{1}{2n} \leq u < \frac{1}{2(n-1)}$, then we can produce a broadband excitation as

\begin{equation}
\label{eq:thirdexcitation}
U_1 = \Delta (\omega, \frac{T}{2}) \ U(\omega, \frac{\pi}{2n}) \ \left ( \Delta (\omega, T) \ U(\omega, \frac{\pi}{2n}) \right )^{n-1}. 
\end{equation}

Thus we can produce broadband excitation for arbitary small rf-amplitude or viceversa, for a given rf-amplitude, we can cover arbitrary large bandwidths.

We talked about broadband excitations. Now we discuss broadband $\frac{\pi}{2}$ rotations. This is simply obtained from above by an initial double sweep.
Thus

\begin{equation}
\label{eq:firstrotation}
U_1 = \Delta (\omega, \frac{T}{2}) \ U(\omega, \frac{\pi}{2}) \ \Delta (\omega, \frac{T}{2}),
\end{equation} is a $\frac{\pi}{2}$ rotation around $x$ axis.

 If peak amplitude is $ \frac{1}{4} \leq u < \frac{1}{2}$ we can prepare $U(\omega, \frac{\pi}{4})$ and combine two such rotations with three double sweeps as follows

\begin{equation}
\label{eq:secondrotation}
U_1 = \Delta (\omega, \frac{T}{2}) \ U(\omega, \frac{\pi}{4}) \ \Delta (\omega, T) \ U(\omega, \frac{\pi}{4})  \ \Delta (\omega, \frac{T}{2}),
\end{equation} to get a broadband $\frac{\pi}{2}$ rotation around $x$ axis.

In general, if $ \frac{1}{2n} \leq u < \frac{1}{2(n-1)}$, then we can produce a broadband $\frac{\pi}{2}$, $x$ rotation as

\begin{equation}
\label{eq:thirdrotation}
U_1 = \Delta (\omega, \frac{T}{2}) \ U(\omega, \frac{\pi}{2n}) \ \left ( \Delta (\omega, T) \ U(\omega, \frac{\pi}{2n}) \right )^{n-1}  \Delta (\omega, \frac{T}{2}).
\end{equation}

\section{Simulations}

\begin{figure}[htb]
\centering
\begin{tabular}{cc}
\includegraphics[scale=.25]{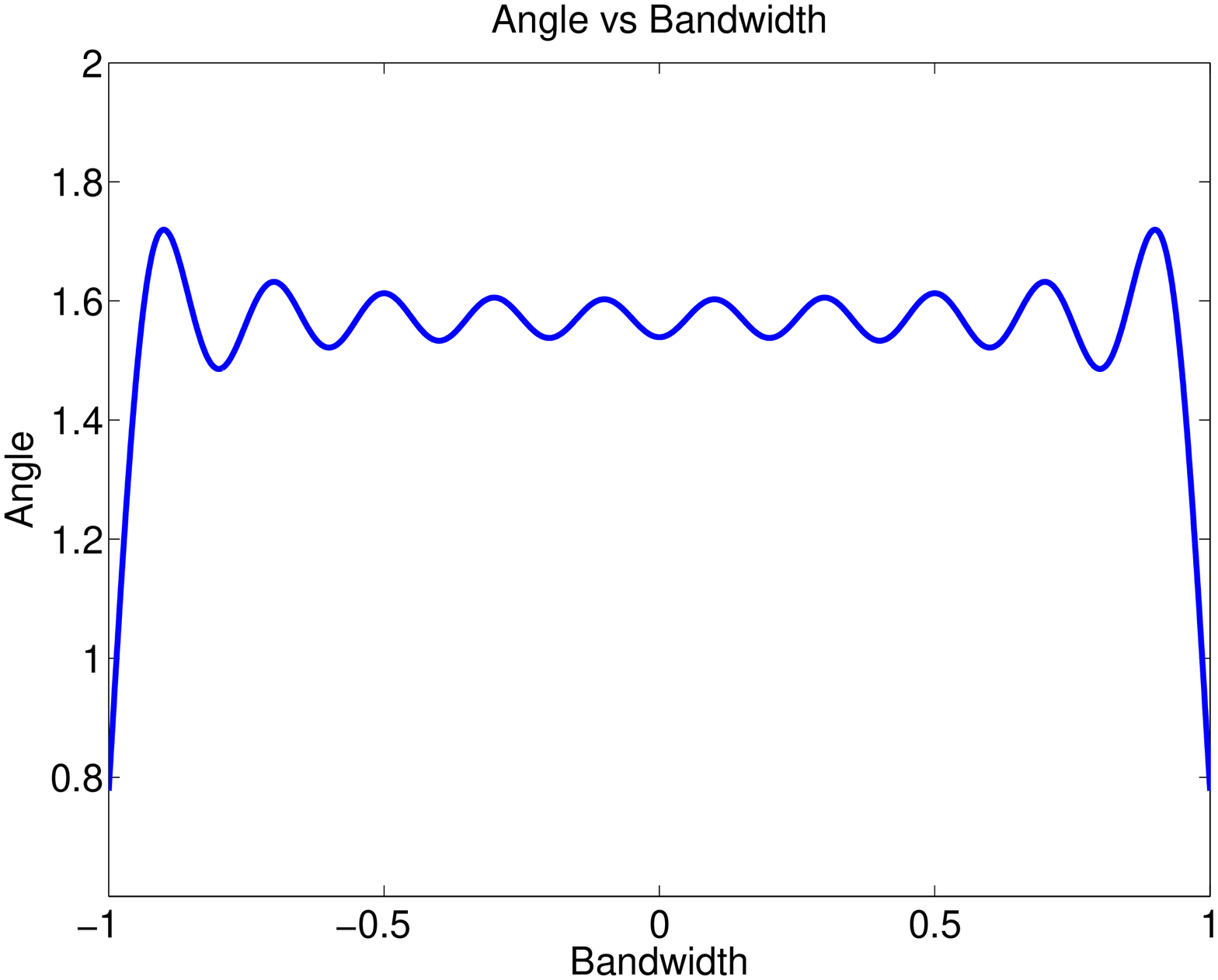} & \includegraphics[scale=.25]{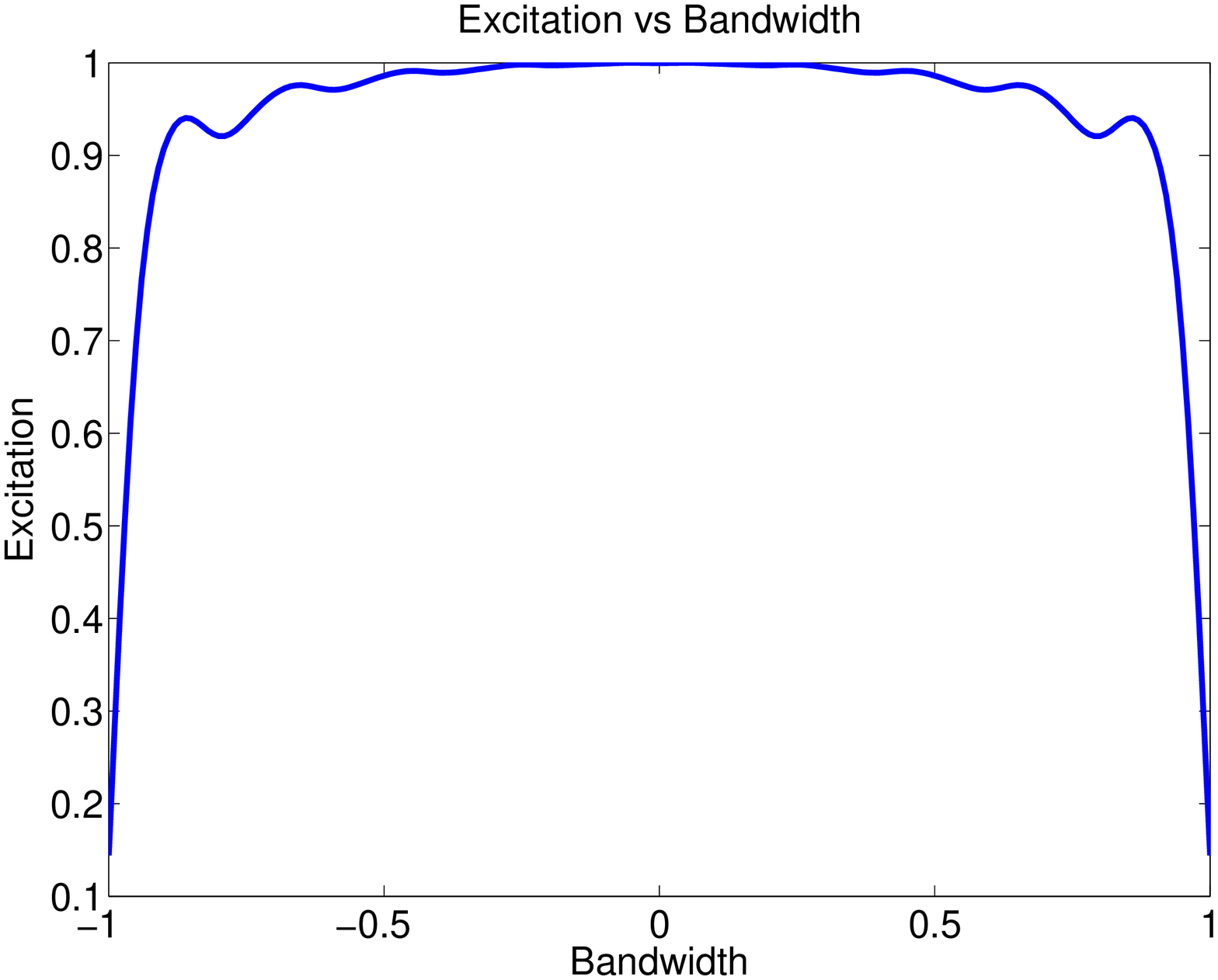}
\end{tabular}
\caption{Left panel shows value of the Eq. (\ref{eq:fourierseries}) as a function of bandwidth when we choose $T = 20 \pi$ and $\Delta = \frac{\pi}{20}$. 
The right panel shows the excitation profile i.e., the $-y$ cordinate of the Bloch vector after application of the pulse in Eq. (\ref{eq:firstexcitation}), with $u_k$ as in Eq. (\ref{eq:fouriercoeff}) and we assume that adiabatic inversion is ideal. The peak rf-amplitude $A \sim \frac{1}{2}$.}\label{fig:fourierangle}
\end{figure}

\begin{figure}[htb]
\centering
\includegraphics[scale=.7]{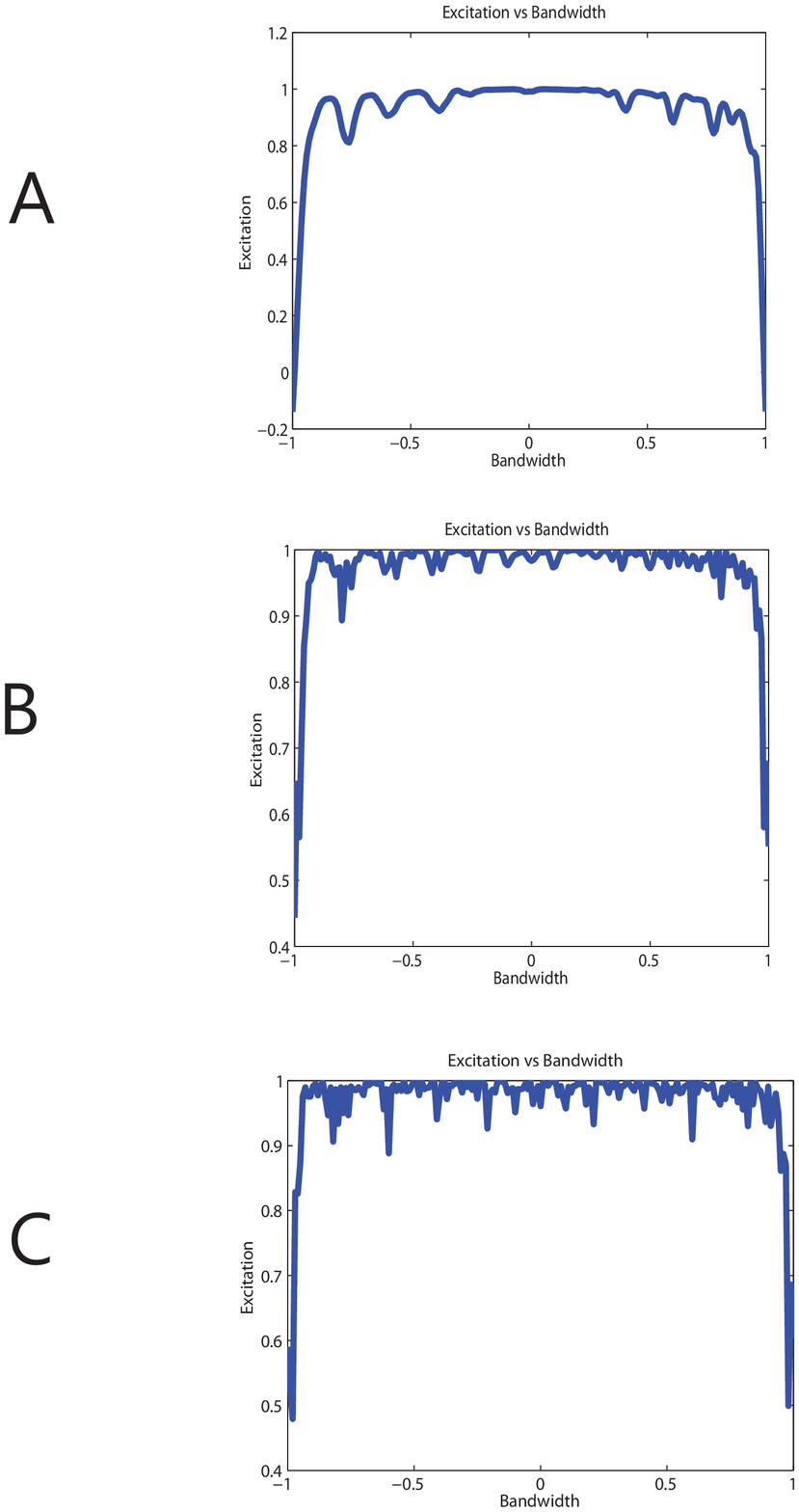} 
\caption{Fig. A shows the excitation profile (the $-y$ cordinate of Bloch vector) for the basic excitation pulse in Eq. (\ref{eq:firstexcitation}) with peak amplitude $A = \frac{1}{2}$.
Fig. B shows the excitation profile for the excitation pulse in Eq. (\ref{eq:secondexcitation}) with peak amplitude $A = \frac{1}{4}$.
Fig. C shows the excitation profile for the basic excitation pulse in Eq. (\ref{eq:thirdexcitation}) with $n=3$,  with peak amplitude $A = \frac{1}{6}$.}\label{fig:excitation}
\end{figure}

\begin{figure}[htb]
\centering
\includegraphics[scale=.7]{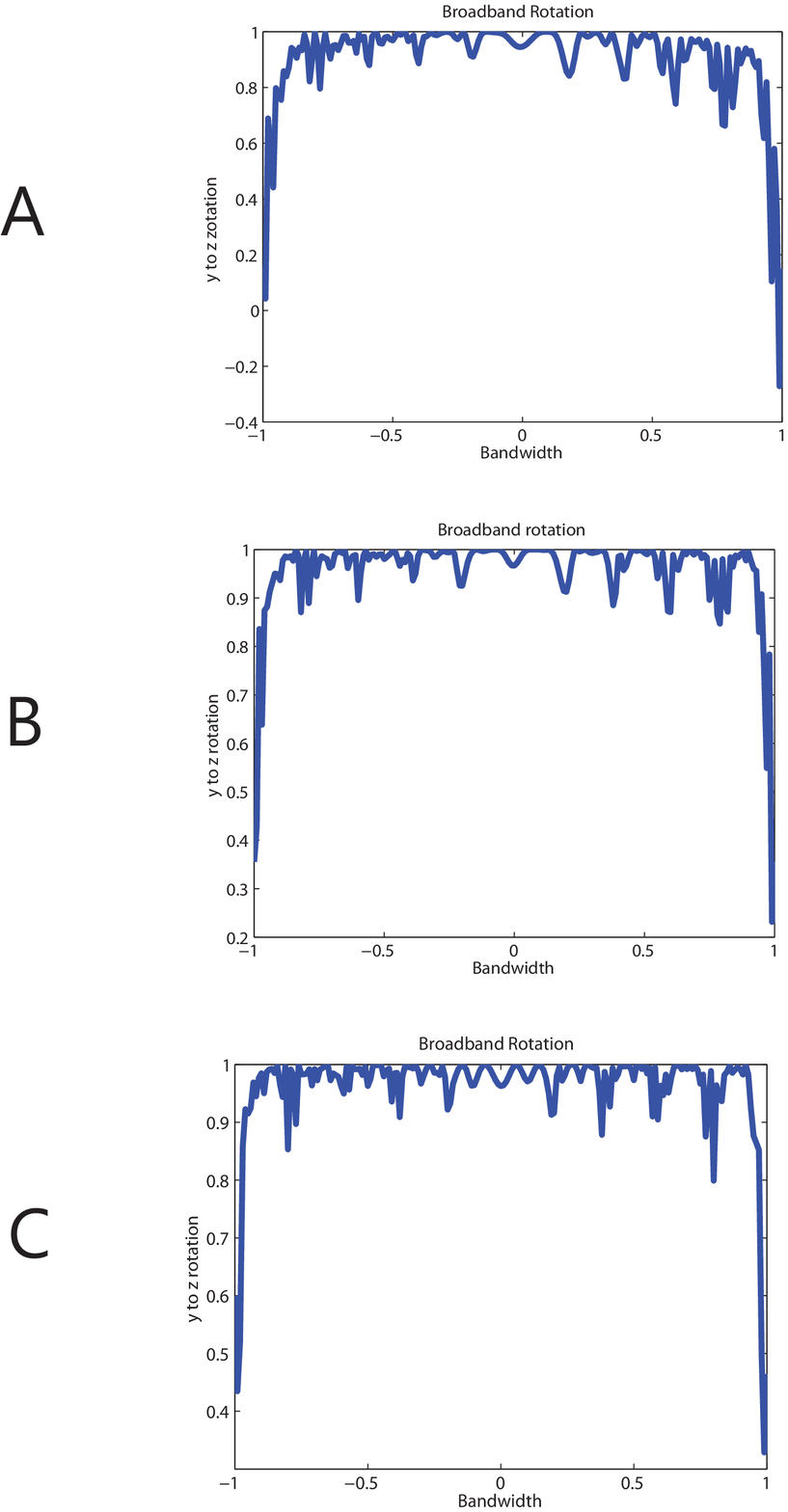}
\caption{Fig. A shows the $y$ to $z$ rotation profile (the $z$ cordinate of Bloch vector) for the broadband $x$ rotation pulse in Eq. (\ref{eq:firstrotation}) with peak amplitude $A = \frac{1}{2}$. Fig. B shows the rotation profile for the  broadband $x$ rotation pulse in Eq. (\ref{eq:secondrotation}) with peak amplitude $A = \frac{1}{4}$.
Fig. C shows the rotation profile for the  broadband $x$ rotation pulse in Eq. (\ref{eq:thirdrotation}) with $n=3$,  with peak amplitude $A = \frac{1}{6}$.}\label{fig:rotation}
\end{figure}

We normalize $\omega$ in Eq. (\ref{eq:basiceq}), to take values in the range $[-1, 1]$. We choose time $\frac{T}{2} = M \pi$, where we choose $M = 10$ and $N = 20$ in 
$\Delta t = \frac{\pi}{N}$ in Eq. (\ref{eq:fourierapprox}). Choosing $\theta = \frac{\pi}{2}$ and coeffecients $u_k$ as in Eq. (\ref{eq:fouriercoeff}), we get the value of the
Eq. (\ref{eq:fourierseries}) as a function of bandwidth as shown in left panel of Fig. \ref{fig:fourierangle}. This is a decent approximation to $\frac{\pi}{2}$ over the entire bandwidth. The right panel of Fig. \ref{fig:fourierangle}, shows the excitation profile i.e., the $-y$ cordinate of the bloch vector after application of the pulse in Eq. (\ref{eq:firstexcitation}), where we assume that adiabatic inversion is ideal. The peak rf-amplitude $A \sim \frac{1}{2}$.

Next, we implement the nonideal adiabatic sweep, by sweeping from $[-5, 5]$ in $300$ units of time. This gives a sweep rate $\frac{1}{30} \ll A^2$, where $A = \frac{1}{2}$. The resulting excitation profile of Eq. (\ref{eq:firstexcitation}) is shown in  Fig. \ref{fig:excitation} A, where we show the $-y$ coordinate of the Bloch vector.
For $A = 10$ kHz, this pulse takes $5.5225$ ms, and excites a bandwidth of $[-20, 20]$ kHz.

Next, we simulate the excitation with reduced amplitude $A = \frac{1}{4}$, as in Eq. (\ref{eq:secondexcitation}). This requires to perform double sweep twice, as in  Eq. (\ref{eq:secondexcitation}). Adiabatic sweep is implemented by sweeping $[-5, 5]$ in $1000$ units of time. This gives a sweep rate $\frac{1}{100} \ll A^2$, where $A = \frac{1}{4}$. The resulting excitation profile of Eq. (\ref{eq:secondexcitation}) is shown in  Fig. \ref{fig:excitation} B, where we show the $-y$ coordinate of the Bloch vector.
For $A = 10$ kHz, this pulse takes $16.79$ ms, and excites a bandwidth of $[-40, 40]$ kHz.

Next, we simulate the excitation with reduced amplitude $A = \frac{1}{6}$, as in Eq. (\ref{eq:thirdexcitation}) for $n=3$. This requires to perform double sweep thrice as in  Eq. (\ref{eq:thirdexcitation}). Adiabatic sweep is implemented by sweeping $[-5, 5]$ in $2000$ units of time. This gives a sweep rate $\frac{1}{200} \ll A^2$, where $A = \frac{1}{6}$. The resulting excitation profile of Eq. (\ref{eq:thirdexcitation}) is shown in  Fig. \ref{fig:excitation} C, where we show the $-y$ coordinate of the Bloch vector.
For $A = 10$ kHz, this pulse takes $32.75$ ms, and excites a bandwidth of $[-60, 60]$ kHz.

Next, we simulate the broadband $x$ rotation as in Eq. (\ref{eq:firstrotation}), with peak amplitude $A = \frac{1}{2}$. This requires to perform double sweep twice as in  Eq. (\ref{eq:firstrotation}). Adiabatic sweep is implemented by sweeping $[-5, 5]$ in $1000$ units of time. This gives a sweep rate $\frac{1}{100} \ll A^2$, where $A = \frac{1}{2}$. The resulting excitation profile of Eq. (\ref{eq:firstrotation}) is shown in  Fig. \ref{fig:rotation} A, where we show the $z$ coordinate of the Bloch vector starting from initial $y = 1$.
For $A = 10$ kHz, this pulse takes $32.83$ ms, and excites a bandwidth of $[-20, 20]$ kHz.

Next, we simulate the broadband $x$ rotation as in Eq. (\ref{eq:secondrotation}), with peak amplitude $A = \frac{1}{4}$. This requires to perform double sweep thrice as in  Eq. (\ref{eq:secondrotation}). Adiabatic sweep is implemented by sweeping $[-5, 5]$ in $1200$ units of time. This gives a sweep rate $\frac{1}{120} \ll A^2$, where $A = \frac{1}{4}$. The resulting excitation profile of Eq. (\ref{eq:secondrotation}) is shown in  Fig. \ref{fig:rotation} B, where we show the $z$ coordinate of the Bloch vector starting from initial $y = 1$.
For $A = 10$ kHz, this pulse takes $29.6462$ ms, and excites a bandwidth of $[-40, 40]$ kHz.

Next, we simulate the broadband $x$ rotation as in Eq. (\ref{eq:thirdrotation}), with peak amplitude $A = \frac{1}{6}$ and $n=3$. This requires to perform double sweep four times as in  Eq. (\ref{eq:thirdrotation}). Adiabatic sweep is implemented by sweeping $[-5, 5]$ in $2400$ units of time. This gives a sweep rate $\frac{1}{240} \ll A^2$, where $A = \frac{1}{6}$. The resulting excitation profile of Eq. (\ref{eq:thirdrotation}) is shown in  Fig. \ref{fig:rotation} C, where we show the $z$ coordinate of the Bloch vector starting from initial $y = 1$.
For $A = 10$ kHz, this pulse takes $51.9267$ ms, and excites a bandwidth of $[-60, 60]$ kHz.

\section{Experimental}

\begin{figure}[htb]
\centering
\includegraphics[scale=.5]{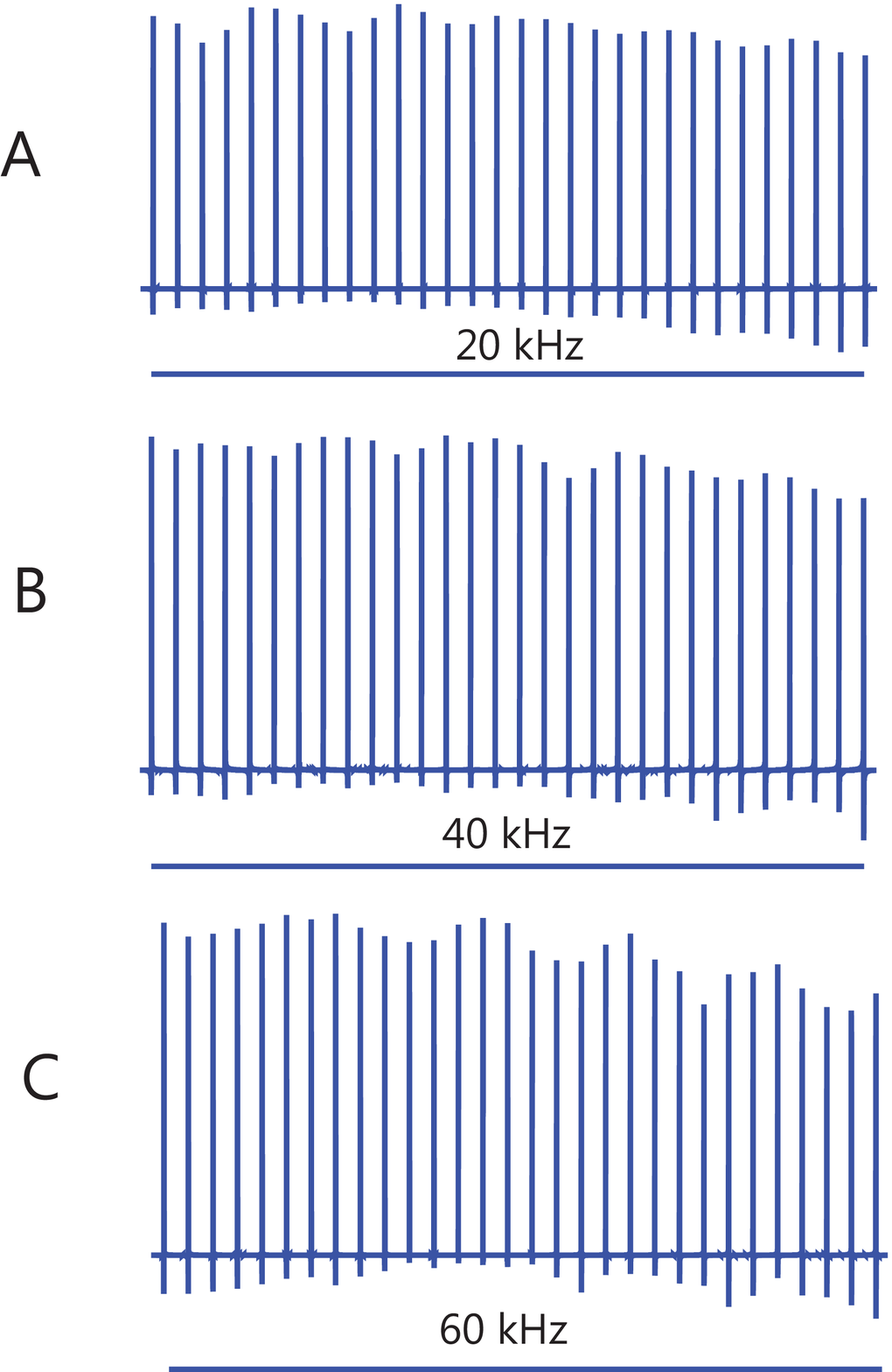}
\caption{Fig. A, B, C show the experimental excitation profile of pulse sequences in Fig. \ref{fig:excitation}A,  \ref{fig:excitation}B and   \ref{fig:excitation}C, respectively, 
in a sample of $99.5\%$ D$_2$0. The offset is varied over the range as shown and the peak rf power of all pulses is 10 kHz. The duration of the pulses is $5.5225$, $16.79$ and 
$32.7467$ ms respectively. }\label{fig:dsweep}
\end{figure}

All experiments were performed on a 750 MHz (proton
frequency) NMR spectrometer at 298 K.
Fig. \ref{fig:dsweep} shows the experimental excitation profiles for the residual HDO signal in a sample
of $99.5\%$ D$_2$0 displayed as a function of resonance offset. Fig.  \ref{fig:dsweep}A shows the excitation profile of broadband excitation sequence in   Fig. \ref{fig:excitation} A.  The peak amplitude of 
the rf-field is 10 kHz and duration of the pulse is 5.5225 ms. The pulse sequence uses one double sweep. The offset is varied over a range of 20 kHz with on-resonance at 3.53 kHz (4.71 ppm). 
Theoretically the pulse covers a bandwidth of 40 kHz. We only show its performance over a 20 kHz range.

Fig.  \ref{fig:dsweep}B shows the excitation profile of the broadband excitation sequence in   Fig. \ref{fig:excitation} B.  The peak amplitude of 
the rf-field is 10 kHz and duration of the pulse is 16.79 ms. The pulse sequence uses two double sweeps. The offset is varied over a range of 40 kHz. Theoretically the pulse covers a bandwidth of 80 kHz. We only show its performance over a 40 kHz range.

Fig.  \ref{fig:dsweep}C shows the excitation profile of the broadband excitation sequence in   Fig. \ref{fig:excitation} C.  The peak amplitude of 
the rf-field is 10 kHz and duration of the pulse is 32.7467 ms. The pulse sequence uses three double sweeps. The offset is varied over a range of 60 kHz. Theoretically the pulse covers a bandwidth of 120 kHz. We only show its performance over a 60 kHz range.

\begin{figure}[htb]
\centering
\includegraphics[scale=.5]{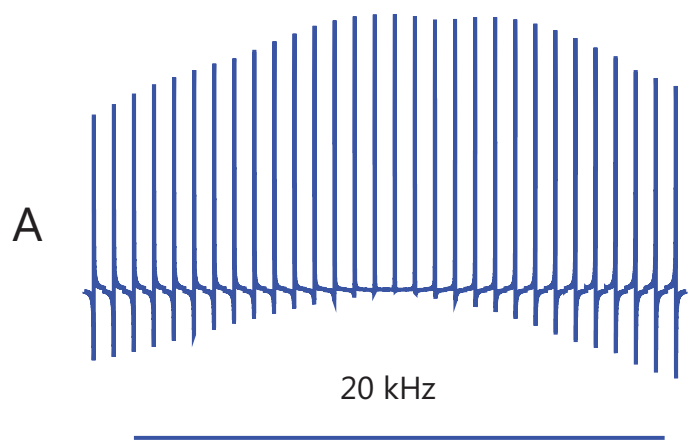}
\caption{Fig. A show the experimental excitation profile of a $10$ kHz, $25 \mu s$, $90^{\circ}$ hard pulse in a sample of $99.5\%$ D$_2$0. The offset is varied over the range as shown. }\label{fig:90scan}
\end{figure}

For comparison we plot in Fig. \ref{fig:90scan},  the excitation profile of a $10$ kHz, $25 \mu s$, $90^{\circ}$ hard pulse in a sample of $99.5\%$ D$_2$0. The offset is varied over the range of $20$ kHz.

\begin{figure}[htb]
\centering
\includegraphics[scale=.5]{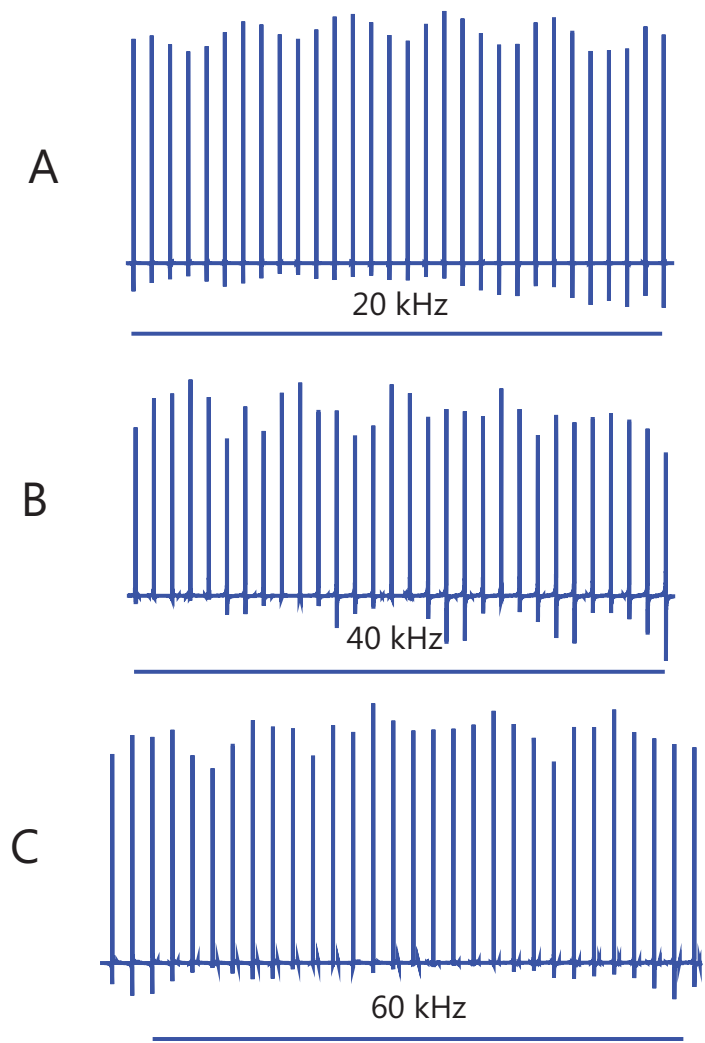}
\caption{Fig. A, B, C show the experimental excitation profile of pulse sequences in Fig. \ref{fig:rotation}A,  \ref{fig:rotation}B and  \ref{fig:rotation}C respectively, 
in a sample of $99.5\%$ D$_2$0. The offset is varied over the range as shown and the peak rf power of all pulses is 10 kHz. The duration of the pulses is $32.83$, and 
$29.6462$ and $51.93$ ms respectively.}\label{fig:brot}
\end{figure}

Fig.  \ref{fig:brot}A shows the excitation profile of the 
broadband rotation sequence in   Fig. \ref{fig:rotation} A.  The peak amplitude of 
the rf-field is 10 kHz and duration of the pulse is 32.83 ms. The sequence uses two double sweeps. The offset is varied over a range of 20 kHz. 
Theoretically the pulse covers a bandwidth of 40 kHz. We only show its performance over a 20 kHz range.

Fig.  \ref{fig:brot}B shows the excitation profile of the 
broadband rotation sequence in   Fig. \ref{fig:rotation} B as an excitation pulse.  The peak amplitude of 
the rf-field is 10 kHz and duration of the pulse is 29.6462 ms. The sequence uses three double sweeps. The offset is varied over a range of 40 kHz. 
Theoretically the pulse covers a bandwidth of 80 kHz. We only show its performance over a 40 kHz range.

Fig.  \ref{fig:brot}C shows the excitation profile of the 
broadband rotation sequence in   Fig. \ref{fig:rotation} C as an excitation pulse.  The peak amplitude of 
the rf-field is 10 kHz and duration of the pulse is 51.93 ms. The sequence uses four double sweeps. The offset is varied over a range of 60 kHz. 
Theoretically the pulse covers a bandwidth of 120 kHz. We only show its performance over a 60 kHz range.

\begin{figure}[htb]
\centering
\includegraphics[scale=.5]{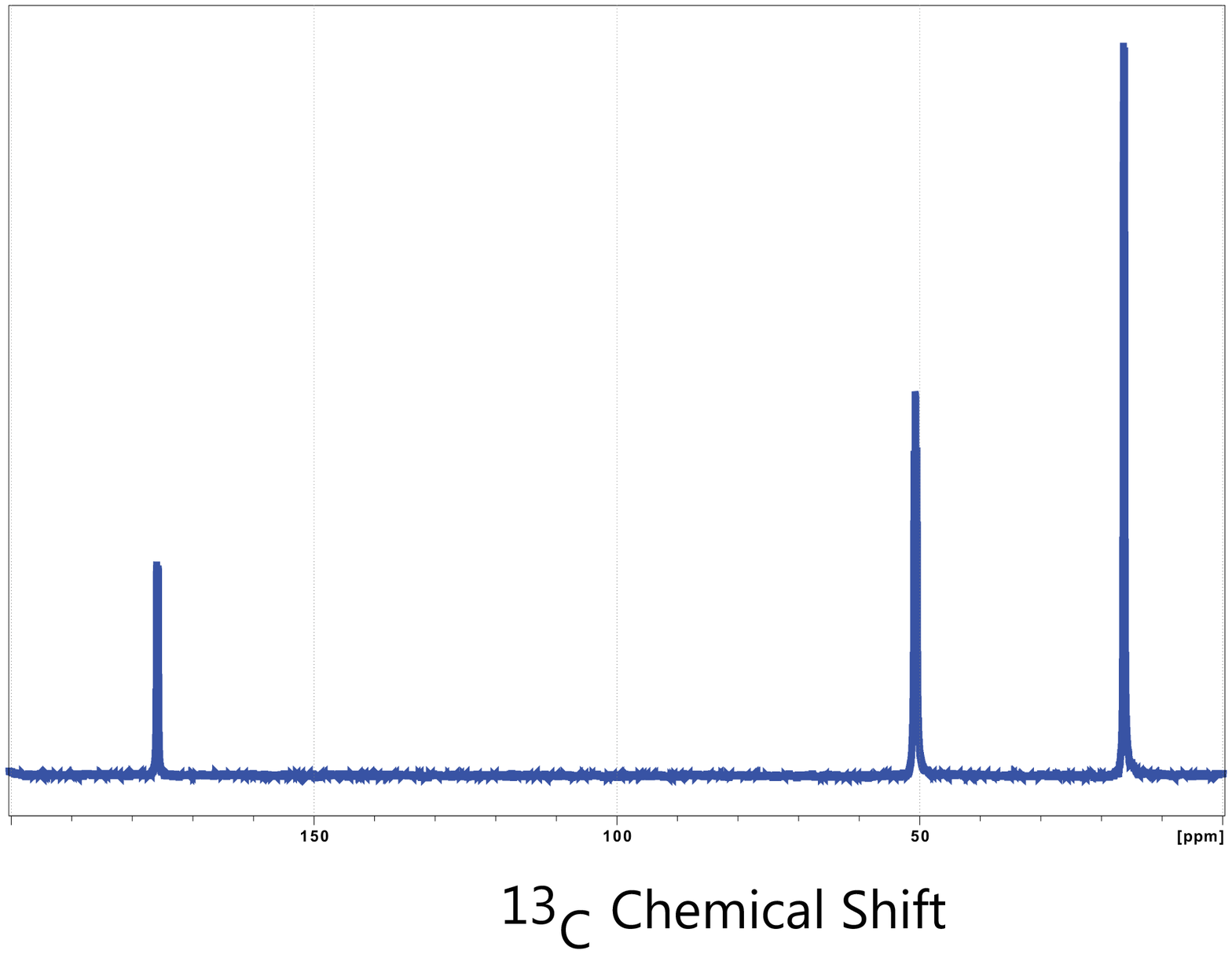}
\caption{Fig. shows excitation of $^{13}$C in sample of labelled Alanine. The excitation pulse is a $1.75$ ms pulse as in Fig. \ref{fig:pulsesequence}. The inversion is done with a $500 \mu s$ chirp pulse and delay $T = 500 \mu s$. The rf-pulses have a peak amplitude $10$ kHz.}\label{fig:cpr}  
\end{figure}

Fig. \ref{fig:cpr} shows the excitation pulse with only one double sweep, as in Fig. \ref{fig:pulsesequence}, applied on a $^{13}$C sample of labelled Alanine. The excitation pulse is a $1.75$ ms pulse.  The inversion is done with a $500 \mu s$, $60$ kHz sweep width, chirp pulse available in Bruker shaped pulse libray and delay $T = 500 \mu s$ in Fig. \ref{fig:pulsesequence}. The rf-pulses in Fig. \ref{fig:pulsesequence} have a peak amplitude $10$ kHz and bandwidth of the pulse is $[-20, 20]$ kHz.

\section{Conclusion}

In this paper we showed design of broadband excitation and rotation pulses. We first showed how by use of Fourier series, we can design a pulse that does broadband excitation to the equator of Bloch sphere. The phase of excitation is linearly dispered as function of offset, which is refoucsed by nesting free evolution between adiabatic inversion pulses. We then showed we can extend the design to arbitrary large bandwidths without increasing peak rf-amplitude. Finally, we extented the method to produce broadband rotations. The pulse duration of the pulse sequences is largely limited by time of adiabatic sweeps. This increases, if we have to excite larger bandwidths. Both because we need more adiabatic inversions and also more time is needed to do adaiabatic inversion over a larger Bandwidth. However adiabatic sweeps is not the only method availible to do broadband inversion. We can use tecniques like optimal control \cite{kobzar} or multiple rotating frames \cite{ultrabroadband} or other methods to do these inversions much faster and hence we can reduce the time of the proposed pulse sequences.
The principle merit of the proposed pulse sequences is the analytical tractability and conceptual simplicity of the design.

\section{Acknowledgement}
The authors would like to thank the HFNMR lab facility at IIT Bombay, funded by RIFC, IRCC, where the data was collected.

\end{document}